\documentclass[aps,pra,preprint,showpacs]{revtex4}
\usepackage{dcolumn}                 
\newcolumntype{w}[1]{D{.}{.}{#1}}
\newcommand*{\centt}[1]{\multicolumn{1}{c}{#1}}
\newcommand*{\cent}[1]{\multicolumn{1}{c}{$#1$}}
\usepackage{times}
\usepackage{nicefrac}
\usepackage{amsmath}
\usepackage{amsfonts}
\usepackage{amssymb}
\usepackage{amsthm}

\begin{document}
\preprint{Version 3.0}

\title{Relativistic, QED, and finite nuclear mass
       corrections for low-lying states of Li and Be$^+$}

\author{Mariusz Puchalski}
\email[]{mpuchals@fuw.edu.pl}

\author{Krzysztof Pachucki}
\email[]{krp@fuw.edu.pl}

\affiliation{Institute of Theoretical Physics, University of Warsaw,
             Ho\.{z}a 69, 00-681 Warsaw, Poland}

\begin{abstract}
Accurate results for nonrelativistic energy, relativistic, QED, and finite
nuclear mass corrections are obtained for $2^1S_{1/2}$, $3^1S_{1/2}$
and $2^1P_{1/2}$ states of the Li atom and Be$^+$ ion. Our computational approach
uses the Hylleraas basis set with the analytic integration and
recursion relations. From comparison of experimental results
for the isotope shifts to theoretical predictions 
including nuclear polarizabilities,
we obtain nuclear charge radii for Li and Be isotopes.

\end{abstract}

\maketitle
\section{Introduction}
The accurate evaluation of energy levels in atoms and ions
requires inclusion of both the electron correlations and
quantum electrodynamic (QED) effects. At present, in spite of
significant theoretical effort \cite{lindgren} there is no universal
computational method which treats accurately  correlations and QED effects all
together.  For heavy few electron systems, electron interactions can be
treated perturbatively within systematic QED approach \cite{shab}.
For light systems involving few electrons, the most fundamental approach 
is based on the expansion
of energy levels in the fine structure constant $\alpha$, and also in
the electron-nucleus mass ratio. At the same time, electron correlations
are treated accurately by the use of explicitly correlated
basis sets. This method has been advanced significantly in the last years
by the calculation of $m\,\alpha^6$ corrections to helium
energy levels \cite{h6hel}, $m\,\alpha^7$ to helium fine structure
\cite{drake_fs, krp_fs}, and $m\,\alpha^5$ corrections in three- and four-electron 
systems \cite{dy_bethe1, dy_bethe2, beqed}.
The achieved numerical precision is sufficient
to determine nuclear properties from isotope shift measurements of transition energies.
For example, the most accurate determination of the deuteron charge radius
comes from the measurement of the 1S-2S transition in hydrogen and deuterium \cite{hd_iso}.
Recently a series of measurements of isotope shifts in helium \cite{he6_exp,he8_exp}
and lithium isotopes \cite{lit_iso1, lit_iso2}
together with the intensive calculations of theoretical energy levels
\cite{dy_bethe2, dy_li_iso1, dy_li_iso2, mp_li_iso}
brought the accurate values of charge radii of short-lived nuclei
with respect to the stable isotope. However, at present,
theoretical methods are not capable of predicting energy levels
with such an accuracy, which would make available the absolute
determination of nuclear charge radii, except for the hydrogenic systems,
where accuracy is limited only by small $m\,\alpha^2\,(Z\,\alpha)^6$
higher order two-loop corrections \cite{yerokhin2}.

The most important in the accurate calculation of (light) atomic energy levels
is the precise representation of the nonrelativistic wave function.
The frequently used explicitly correlated Gaussian (ECG) functions
give very accurate nonrelativistic energies, but the wave function
does not satisfy the cusp condition, and for this reason this representation
cannot be used for the calculation of $m\,\alpha^6$ and higher order
corrections. Also, the estimation of numerical uncertainties within the ECG
method is quite problematic.
More difficult to use is the Hylleraas basis set \cite{king_lit,yan_lit, rec1},
but its achieved accuracy exceeds significantly results with ECG for 
three electron systems. In order to solve accurately Schr\"odinger equation
we use the Hylleraas basis set with the number of functions of about
$10\,000$ for S-states and $14\,000$ for P-states.
The calculations of matrix elements of nonrelativistic Hamiltonian,
as well as relativistic and QED operators are performed analytically using
newly developed recursion relations for the Hylleraas integrals
\cite{rec1, rec2, li_ground, rec3}.

In this work we present the most accurate
calculations of nonrelativistic, leading relativistic and QED contributions
including finite nuclear mass corrections and nuclear polarizability,
to energy levels of $2^1S_{1/2}$, $3^1S_{1/2}$, and $2^1P_{1/2}$ states of Li and Be$^+$.
As a main result, we obtain ionization energies of Li and Be$^+$,
transition energies $2^1S_{1/2} - 3^1S_{1/2}$ and $2^1S_{1/2} - 2^1P_{1/2}$,
and corresponding isotope shifts.
In comparison to the former work of Yan {\em et al.} \cite{dy_bethe2}
our results are in agreement for transition frequencies, but in slight 
disagreement for the isotope shifts in Be$^+$. For the Li isotope shift
both Ref. \cite{dy_bethe2} and this work are in agreement with \cite{mp_li_iso}.
In all cases our numerical precision is about an order of magnitude higher, 
with the total uncertainty dominated 
by higher order terms. The comparison to experimental values
\cite{sansonetti, bushaw1, bushaw2, ralchenko}, apart from agreement
with experimental transition energies in both Li and Be$^+$,
reveals small discrepancy for the Be$^+$ ionization energy with
the NIST data \cite{ralchenko}.
Together with these benchmarking results, from our theoretical predictions
and measured isotope shifts \cite{lit_iso1, lit_iso2, be_iso},
we obtain improved nuclear charge radii for various isotopes
including halo nuclei $^{11}$Li and $^{11}$Be.

\section{Relativistic and QED corrections with the nuclear charge radius}
The energy level $E(\alpha,\eta)$ as a function of $\alpha$ and $\eta=-\mu/m_N
= -m/(m+m_N)$
is expanded in power series in its arguments
\begin{eqnarray}
E(\alpha,\eta) &=&
m\,\alpha^2\,\bigl[{\cal E}^{(2,0)}+\eta\,{\cal E}^{(2,1)}+\eta^2\,{\cal E}^{(2,2)}\bigr]
+m\,\alpha^4\,\bigl[{\cal E}^{(4,0)}+\eta\,{\cal E}^{(4,1)}\bigr]\nonumber \\ &&
+m\,\alpha^5\,\bigl[{\cal E}^{(5,0)}+\eta\,{\cal E}^{(5,1)}\bigr]
+m\,\alpha^6\,\bigl[{\cal E}^{(6,0)}+\eta\,{\cal E}^{(6,1)}\bigr]
+m\,\alpha^7\,{\cal E}^{(7,0)},
\end{eqnarray}
and each coefficient is calculated separately from the expectation value
of the corresponding Hamiltonian or the operator.
The leading terms ${\cal E}^{(2,0)}$, ${\cal E}^{(2,1)}$, and ${\cal E}^{(2,2)}$
result from nonrelativistic Hamiltonian for three electrons and the nucleus with
Coulomb interactions between them,
\begin{equation}
H_0 = \sum_a\frac{p_a^2}{m} + \frac{p_N^2}{m_N}-\sum_a\frac{Z\,\alpha}{r_a}
+\sum_{a>b} \frac{\alpha}{r_{ab}}. \label{h0}
\end{equation} 
In order to calculate relativistic and nuclear recoil corrections
for atomic systems with the finite size nucleus,
one should at first properly define the nuclear charge radius. Therefore,
let us consider the interaction of the particle having spin $\vec s$
and the charge $e$ with the electric field.
The leading interaction related to the finite size is
\begin{equation}
\delta H = -\frac{e}{6}\,\biggl(\langle r_{\rm ch}^2\rangle\,\delta^{ij}+
(s^i\,s^j)^{(2)}\,Q_E\biggr)\,\partial^j E^i, \label{01}
\end{equation}
where $\langle r_{\rm ch}^2\rangle$ is the averaged square of the charge radius,
$Q_E$ is related to the electric quadrupole moment $Q$ by $Q_E =
3/(s_N\,(2\,s_N-1))\,Q$, and
\begin{equation}
(s^i\,s^j)^{(2)} = \frac{1}{2}\,s^i\,s^j + \frac{1}{2}\,s^j\,s^i -
\frac{\delta^{ij}}{3}\,\vec s^{\,2}.
\end{equation}
For the point spin $s=1/2$ particle $\langle r_{\rm ch}^2\rangle$ does not vanish
and is equal to $3/(4\,m^2)$, it is the so-called Darwin term and it
depends on the value of the spin $s$ \cite{darwin}.
This term is the source of ambiguity in the definition of nuclear charge
radii \cite{friar}. If we want the point
particle to have the vanishing charge radius, the Darwin term should be
excluded from $\langle r_{\rm ch}^2\rangle$. However,
the value of the Darwin term for the arbitrary spin point
particle with the arbitrary magnetic moment is unknown.
Therefore, we propose the universal definition of $\langle r_{\rm ch}^2\rangle$ by Eq. (\ref{01}),
and thus include the Darwin term within the charge radius.
A similar, but even more complicated problem appears when QED effects
are being included. Now, assuming the definition of charge radius
by Eq. (\ref{01}), the atomic Hamiltonian including relativistic
corrections and neglecting magnetic moment anomaly for electrons
is of the form \cite{bs}
\begin{eqnarray}
H_{BP} &=& \sum_a H_a + \sum_{a>b} H_{ab}+ H_N +\sum_{a} H_{aN}, \label{02}\\
H_a &=& \frac{\vec p_a^2}{2\,m}-\frac{\vec p^4_a}{8\,m^3},\label{03}
\\
H_{ab} &=& \alpha\,\biggl\{
\frac{1}{r_{ab}} -\frac{1}{2\,m^2}\, p_a^i\,
\biggl(\frac{\delta^{ij}}{r_{ab}}+\frac{r^i_{ab}\,r^j_{ab}}{r^3_{ab}}
\biggr)\, p_b^j + \frac{\pi}{m^2}\, \delta^3(r_{ab})
+\frac{1}{m^2}\,\frac{s_a^i\,s_b^j}{r_{ab}^3}\,
\biggl(\delta^{ij}-3\,\frac{r_{ab}^i\,r_{ab}^j}{r_{ab}^2}\biggr)
\nonumber \\ &&
+\frac{1}{2\,m^2\,r_{ab}^3} \biggl[
2\,\vec s_a\cdot\vec r_{ab}\times\vec p_b -
2\,\vec s_b\cdot\vec r_{ab}\times\vec p_a +
\vec s_b\cdot\vec r_{ab}\times\vec p_b
-\vec s_a\cdot\vec r_{ab}\times\vec p_a\biggr]\biggr\},\label{04}
\\
H_N &=& \frac{\vec p_N^{\,2}}{2\,m_N}-\frac{\vec p^{\,4}_N}{8\,m_N^3},\label{05}
\\
H_{aN} &=& -Z\,\alpha\,\biggl\{
\frac{1}{r_{a}} -\frac{1}{2\,m\,m_N}\, p_a^i\,
\biggl(\frac{\delta^{ij}}{r_{a}}+\frac{r^i_{a}\,r^j_{a}}{r^3_{a}}
\biggr)\, p_N^j -
\frac{2\,\pi}{3}\,\biggl(\langle r_{\rm ch}^2\rangle+\frac{3}{4\,m^2}\biggr) \, \delta^3(r_{a})
\nonumber \\ && -
\frac{4\,\pi\,g_N}{3\,m\,m_N}\,\vec s_a \cdot\vec s_N\,\delta^3(r_{a})
+\frac{g_N}{2\,m\,m_N}\,\frac{s_a^i\,s_N^j}
{r_{a}^3}\,
\biggl(\delta^{ij}-3\,\frac{r_{a}^i\,r_{a}^j}{r_{a}^2}\biggr)
\nonumber \\ &&
+\frac{1}{2\,r_{a}^3} \biggl[
\frac{2}{m\,m_N}\,\vec s_a\cdot\vec r_{a}\times\vec p_N -
\frac{g_N}{m\,m_N}\,\vec s_N\cdot\vec r_{a}\times\vec p_a +
\frac{(g_N-1)}{m_N^2}\,\vec s_N\cdot\vec r_{a}\times\vec p_N
 \nonumber \\ &&
-\frac{1}{m^2}\,\vec s_a\cdot\vec r_{a}\times\vec p_a\biggr]
-\frac{Q_{E}}{6}\,\frac{(s_N^i\,s_N^j)^{(2)}}{r_a^3}\,
\biggl(\delta^{ij}-3\,\frac{r_{a}^i\,r_{a}^j}{r_{a}^2}\biggr)\biggr\}.\label{06}
\end{eqnarray}
In practice
small relativistic terms involving nuclear mass are treated perturbatively
and in this work we neglect all relativistic $O(m/M)^2$ corrections.
These terms become much more important in muonic atoms
and cannot be neglected there.

Let us now consider leading QED corrections of order $m\,\alpha^5$,
which  also include the inelastic contribution $E_{\rm pol}$
due to the nuclear polarizability. Since we do not consider the hyperfine 
structure, the spin of atomic nucleus can be neglected, and 
the QED correction takes the form \cite{drake_book,simple}
\begin{eqnarray}
E^{(5)} &=&-\frac{4\,Z\,\alpha^2}{3}\,
\biggl(\frac{1}{m}+\frac{Z}{M}\biggr)^2\,
\Bigl\langle\sum_a\delta^3(r_a)\Bigr\rangle\,\ln k_0
+\sum_a \langle H_{aN}^{(5)}\rangle
+ \sum_{a>b} \langle H_{ab}^{(5)} \rangle+E_{\rm pol}, \label{07}\\
H_{aN}^{(5)} &=& \frac{Z\,\alpha^2}{2\,\pi\,m^2\,r_{a}^3} \biggl[
\vec s_a\cdot\vec r_{a}\times\vec p_a
-\frac{m}{M}\,\vec s_a\cdot\vec r_{a}\times\vec p_N
\biggr]
+\left[\frac{19}{30}+\ln(\alpha^{-2})\right]\,
\frac{4\,\alpha^2\,Z}{3\,m^2}\,\delta^3(r_a)
 \nonumber \\ &&
+\biggl[\frac{62}{3}+\ln(\alpha^{-2})\biggr]\,
\frac{(Z\,\alpha)^2}{3\,m\,M}\,\delta^3(r_a)
-\frac{7}{6\,\pi}\,\frac{m^2}{M}\,(Z\,\alpha)^5\,
P\left[\frac{1}{(m\,\alpha\,r_{a})^3}\right]
\nonumber \\ &&
+\frac{4}{3}\,\frac{Z^3\,\alpha^2}{M^2}\,
\ln\biggl(\frac{M}{m\,\alpha^2}\biggr)\,\delta^3(r_a),\label{09}
\\
H_{ab}^{(5)} &=& \frac{\alpha^2}{\pi\,m^2}\,\biggl[
\frac{s_a^i\,s_b^j}{r_{ab}^3}\,
\biggl(\delta^{ij}-3\,\frac{r_{ab}^i\,r_{ab}^j}{r_{ab}^2}\biggr)
-\frac{1}{2\,r_{ab}^3}\,
(\vec s_a+\vec s_b)\cdot\vec r_{ab}\times(\vec p_a -\vec p_b)\biggr]
 \nonumber \\ &&
+\frac{\alpha^2}{m^2}\,\left[\frac{164}{15}+\frac{14}{3}\,\ln\alpha
\right]\,\delta^3(r_{ab})
-\frac{7}{6\,\pi}\,m\,\alpha^5\,P\left[\frac{1}{(m\,\alpha\,r_{ab})^3}\right],
\label{10}
\end{eqnarray}
where
\begin{eqnarray}
\ln k_0
&\equiv&\frac{\left\langle \sum_a\vec{p}_a \,(H_0-E_0)\,
\ln\bigl[\frac{2\,(H_0-E_0)}{\alpha^2\,m}\bigr]\,
\sum_b\vec{p}_b \right\rangle}{2\,\pi\,\alpha\,Z\,
\Bigl\langle\sum_c\delta^3(r_c)\Bigr\rangle}\,,\label{08}\\
\langle\phi| P\left[\frac{1}{r^3}\right]|\psi\rangle &=&
\lim_{a\rightarrow 0}\int {\rm d}^3 r\,
\phi^*(\vec r)\left[\frac{1}{r^3}\,\Theta(r-a)
+ 4\,\pi\,\delta^3(r)\,
(\gamma+\ln a)\right]\,\psi(\vec r)\,. \label{08p}
\end{eqnarray}
The electron--electron terms have been simplified in the above,
since $\delta^3(r_{ab})$ does not vanish only for singlet states,
therefore $\vec{s}_a\cdot\vec{s}_b\,\delta^3(r_{ab}) =
-3/4 \,\delta^3(r_{ab})$.
We included in $H_{aN}$  the leading logarithmic contribution
that comes from the nuclear self-energy, but neglected all nonlogarithmic
$(m/M)^2$ terms which are proportional to $\delta^3(r_a)$.
They are not known for a general nucleus since they depend
on the nuclear spin and charge distribution within the nucleus.
For the calculation of the isotope shift in Li and Be$^+$
we again neglect all $O(m/M)^2$ terms. These terms are important
for muonic atoms, but their calculation requires proper definition
of the nuclear charge radius including QED effects, and its relation to
the charge radius obtained from a different type of measurements
such as the elastic electron scattering off nuclei.

The last term in Eq. (\ref{07}), $E_{\rm pol}$ is the nuclear polarizability
correction. It is significant for halo nuclei such as $^{11}$Li \cite{mp_li_iso},
or whenever the isotope shift transition in the optical range reaches
sub MHz precision, for example in the 1S-2S transition in deuterium \cite{hd_iso}.
The nuclear polarizability correction is expected to be 
significant also for $^{11}$Be, as this nucleus has the largest known 
$B(E1)$ line strength among all nuclei. For this reason, we calculate it
using experimental \cite{palit} and theoretical \cite{typel}
data for the electric dipole excitation of the $^{11}$Be nucleus.

Considering $m\,\alpha^6$ corrections, they are well known
for the hydrogen.
Results for few-electron atoms are expressed in terms of the effective
Hamiltonian \cite{fw}. Corresponding calculations have been performed only
for low-lying states of helium \cite{h6hel}.
Calculations  for three-electron systems are at present
too difficult  and therefore we use an approximate formula
on the basis of hydrogenic values,
\begin{eqnarray}
E^{(6)} &=& \biggl\{\frac{Z^2\,\alpha^3}{m^2}\,
            \left[\frac{427}{96} - 2 \ln(2) \right]
            +\frac{Z^2\,\alpha^3}{m\,m_N}\,
            \left[\frac{35}{36} - \frac{448}{27 \pi^2} - 2 \ln(2)
            +\frac{6 \zeta(3)}{\pi^2} \right]
\nonumber \\ &&
            +\frac{Z^3\,\alpha^3}{m\,m_N}\,
            \left[4 \ln(2) - \frac{7}{2} \right]\biggr\}
            \pi\,\left\langle\sum_a\delta^3(r_a)\right\rangle.
\end{eqnarray}
It includes dominating electron-nucleus one-loop radiative, radiative recoil
and pure recoil corrections \cite{eides}. We neglect electron-electron radiative
corrections and the purely relativistic corrections, as we expect them to be
relatively small, of order 10\%. The relativistic $m\,\alpha^6$ corrections are also very
difficult to calculate. Its neglect is the leading source of uncertainty
in the theoretical predictions for transition frequencies.
Similarly, the nuclear recoil correction $m^2/M\,\alpha^6$, which is
still significant for the isotope shift,
is also estimated on the basis of the known hydrogenic value
in the above formula. This introduced some uncertainty in the determination
of the isotope shift, from which nuclear charge radii
are obtained. 

Due to numerical importance, one calculates approximately
$m\,\alpha^7$ contribution which is known exactly
only for hydrogenic systems \cite{eides}.
\begin{eqnarray}
E^{(7)}_{\rm H}(n) &=& m\,\frac{\alpha}{\pi}\,\frac{(Z\,\alpha)^6}{n^3}\,
              \bigl[A_{60}(n)+\ln(Z\,\alpha)^{-2}\,A_{61}(n)
              +\ln^2(Z\,\alpha)^{-2}\,A_{62}\bigr]
              \nonumber \\ &&
              +m\,\Bigl(\frac{\alpha}{\pi}\Bigr)^2\,
               \frac{(Z\,\alpha)^5}{n^3}\,B_{50}
              +m\,\Bigl(\frac{\alpha}{\pi}\Bigr)^3\,
              \frac{(Z\,\alpha)^4}{n^3}\,C_{40}.
\end{eqnarray}
It includes one-, two-, and three-loop corrections, and values of $A,B$, and
$C$ coefficients may be found in \cite{eides}.
Following Ref. \cite{drake_book} these hydrogenic values of order $m\,\alpha^7$ are
extrapolated to lithium, according to
\begin{equation}
{\cal E}^{(7)}(Z) = \bigl[2\,{\cal E}^{(7)}(1S,Z)
+{\cal E}^{(7)}(nX,Z-2)\bigr]
\,\frac{\langle \delta^3(r_1) + \delta^3(r_2) + \delta^3(r_3)\rangle_{\rm Li}}
{2\,\langle \delta^3(r)\rangle_{1S,Z}
+ \langle\delta^3(r)\rangle_{nX,Z-2}},
\end{equation}
for $X=S$, and for states with higher angular momenta ${\cal E}^{(7)}(nX,Z)$ is neglected.
We expect this approximate formula to be accurate to 25\%.
This completes QED corrections to transition frequencies and the isotope
shifts in light atomic systems.

\section{Computational method and numerical results}
In the construction of the wave function we closely follow the works of Yan and Drake
in \cite{yan_lit}. The global wave function $\Psi$ for both S and P states
is expressed as a linear combination of $\psi$, the antisymmetrized product ${\cal A}$
of the spatial function $\phi$ and the spin function $\chi$,
\begin{eqnarray}
\psi &=& {\cal A}[\phi(\vec r_1,\vec r_2,\vec r_3)\,\chi]\,,
\label{11}\\
\psi^i_{a} &=& {\cal A}[\phi_a^i(\vec r_1,\vec r_2,\vec r_3)\,\chi]\,,
\label{11p}\\
\phi(\vec r_1, \vec r_2, \vec r_3) &=& e^{-w_1\,r_1-w_2\,r_2-w_3\,r_3}\,
r_{23}^{n_1}\,r_{31}^{n_2}\,r_{12}^{n_3}\,r_{1}^{n_4}\,r_{2}^{n_5}\,r_{3}^{n_6}\,,
\label{12}\\
\phi^i_a(\vec r_1, \vec r_2, \vec r_3) &=& r_a^i\,\phi(\vec r_1, \vec r_2, \vec r_3),
\label{13}\\
\chi &=& \alpha(1)\,\beta(2)\,\alpha(3)-\beta(1)\,\alpha(2)\,\alpha(3)\,,
\label{14}
\end{eqnarray}
with $n_i$ being non-negative integers, $w_i\in {R}_+$, and the subscript $a=1,2,3$.
The matrix element of the nonrelativistic Hamiltonian $H_0$ in Eq. (\ref{h0})
or of any spin independent operator can be expressed after eliminating spin
variables as
\begin{eqnarray}
\langle\psi|H_0|\psi'\rangle &=& \langle
2\,\phi(r_1,r_2,r_3)+2\,\phi(r_2,r_1,r_3)-\phi(r_3,r_1,r_2)\nonumber \\ &&-
\phi(r_2,r_3,r_1)-\phi(r_1,r_3,r_2)-\phi(r_3,r_2,r_1)|
 H_0\,|\phi'(r_1,r_2,r_3)\rangle\,,\label{16}\\
\langle\psi_a^i|H_0|\psi'_b{}^i\rangle &=& \langle
2\,\phi_a^i(r_1,r_2,r_3)+2\,\phi_a^i(r_2,r_1,r_3)-\phi_a^i(r_3,r_1,r_2)\nonumber \\ &&-
\phi_a^i(r_2,r_3,r_1)-\phi_a^i(r_1,r_3,r_2)-\phi_a^i(r_3,r_2,r_1)|
 H_0\,|\phi'_b{}^i(r_1,r_2,r_3)\rangle\,.\label{16p}
\end{eqnarray}
While for S-states Hamiltonian matrix elements can be written in one form,
for P-states with the help of an additional $r_1, r_2, r_3$ permutation
they can take two different forms: $\langle\phi_3^i|H_0|\phi'_3{}^i\rangle$ or
$\langle\phi_2^i|H_0|\phi'_3{}^i\rangle$. Next, all these spatial matrix elements
are expressed as linear combination of Hylleraas integrals,
namely the integrals with respect to $r_i$ of the form
\begin{eqnarray}
f(n_1,n_2,n_3,n_4,n_5,n_6) &=& \int \frac{d^3 r_1}{4\,\pi}\,
                               \int \frac{d^3 r_2}{4\,\pi}\,
                               \int \frac{d^3 r_3}{4\,\pi}\,
                               e^{-w_1\,r_1-w_2\,r_2-w_3\,r_3}
\nonumber \\ &&
r_{23}^{n_1-1}\,r_{31}^{n_2-1}\,r_{12}^{n_3-1}\,r_{1}^{n_4-1}\,r_{2}^{n_5-1}\,r_{3}^{n_6-1}\,,
\label{17}
\end{eqnarray}
with non-negative integers $n_i$. They
are performed analytically for $n_1, n_2,n_3=0,1$ and by recursion
relations for larger $n_i$ using formulas derived in \cite{rec1}.
These recursions give the most accurate numerical values of Hylleraas
integrals among all the methods  developed so far. Nevertheless,
multiple precision arithmetics has to be used in generating the Hamiltonian
matrix, in order to avoid near linear dependence of Hylleraas basis functions.

The total wave function is generated from all $\phi$ in Eq. (\ref{16}) with
$n_i$ satisfying a condition
\begin{equation}
\sum_{i=1}^6 n_i \leq \Omega\,,
\label{18}
\end{equation}
for $\Omega$ between 3 and 12.
For each $\Omega$ we minimize energy with respect to the free parameters $w_i$
in Eq. (\ref{16}). In order to increase the accuracy of the nonrelativistic wave function,
following Yan and Drake \cite{yan_lit},
we divide the whole basis set into five sectors (six sectors for P states), 
each one with its own set of $w_i$'s.
To avoid numerical instabilities, within each sector we drop the terms
with $n_4>n_5$ (or $n_4<n_5$) and for $n_4=n_5$ drop terms with $n_1>n_2$ (or $n_1<n_2$).
This division allows for a significant improvements of nonrelativistic
energies by optimization of all, five for S and six for P states, sets of $w_i$'s.
Numerical results for $2^1S_{1/2}$, $3^1S_{1/2}$, and $2^1P_{1/2}$ of Li and
Be$^+$ for various sizes of basis sets are presented in Table \ref{table1}.
The results for the ground state of Li are in agreement with our previous
evaluation \cite{li_ground}.
Results denoted by $\infty$ are obtained by extrapolation to the infinitely
large (complete) basis set, by fitting the function $X(\Omega) = X_0 +
X_1/\Omega^n $ with some integer $n$. The similar fit is used for all other
matrix elements, presented in the following tables.

Calculation of relativistic corrections, which are given by Eqs. (\ref{02})-(\ref{06}), involve
spin independent and spin dependent terms. The matrix element of spin independent terms
are calculated according to Eq. (\ref{16}), while spin orbit terms are
obtained for $P_{1/2}$ state by using
\begin{eqnarray}
\langle\psi_a|\sum_{c=1}^3\vec Q_c\cdot\vec\sigma_c|\psi_b\rangle_{J=1/2} &=&
i\,\Bigl\langle \vec\phi_a(r_1,r_2,r_3) \Bigl|
-2\,\vec Q_3\times\bigl[\vec\phi_b(r_1,r_2,r_3)+\vec\phi_b(r_2,r_1,r_3)\bigr]
\nonumber\\&&
+(\vec Q_1-\vec Q_2+\vec Q_3)\times\bigl[\vec\phi_b(r_2,r_3,r_1)+\vec\phi_b(r_3,r_2,r_1)\bigr]
\nonumber\\&&
+(-\vec Q_1+\vec Q_2+\vec Q_3)\times\bigl[\vec\phi_b(r_1,r_3,r_2)+\vec\phi_b(r_3,r_1,r_2)\bigr]\Bigr\rangle.
\label{19}
\end{eqnarray}
and the result for $P_{3/2}$ is equal to $-1/2$ of that for $P_{1/2}$.
The tensor spin-spin interaction vanishes for both the $P_{1/2}$ and $P_{3/2}$
states. All these matrix elements
include Hylleraas integrals with $n_i =-1$, which are difficult to obtain accurately.
We use the one-dimensional integral form for $f(-1,0,0;0,0,0)$ and $f(0,0,0;-1,n_5,n_6)$
and other $f'$s with $n_i=-1$ are obtained by recursion relations \cite{rec2,li_ground}.
Since these recursions are not stable numerically,
we used quadruple, sextuple, and octuple precision arithmetics written by
Korobov \cite{korobov} to avoid loss of the numerical precision. 
It was especially important for excited
states. Individual results for various operators are presented in
Table \ref{table2}
and the total relativistic correction in Table \ref{table3}.
By the symbol $[\ldots]_{\rm mp}$ in these tables,
we denote the mass polarization correction, namely
\begin{equation}
[\ldots]_{\rm mp} = 2\,[\ldots]\,\frac{1}{(H-E)'}\sum_{a>b}\,\vec p_a\cdot\vec p_b.
\end{equation}
\begin{widetext}
\squeezetable
\begin{table}[!hbt]
\renewcommand{\arraystretch}{1.0}
\caption{Nonrelativistic energy and relativistic  and finite nuclear mass corrections
  in Li and Be$^+$, $\Omega=\infty$ is a result of extrapolation }
\label{table1}
\begin{ruledtabular}

\begin{tabular}{llllll}
$\Omega$  & \cent{{\cal E}^{(2,0)}}        & \cent{{\cal E}^{(2,1)}}   &
\cent{{\cal E}^{(2,2)}}& \cent{{\cal E}^{(4,0)}}&\cent{{\cal E}^{(4,1)}}\\\hline
\multicolumn{6}{c}{Li $2^1S_{1/2}$} \\
 10       &-7.478\,060\,323\,786\,3 & -7.779\,903\,106\,67 &-1.801\,631\,491 &-12.049\,918\,25 & 10.010\,910\,3\\
 11       &-7.478\,060\,323\,861\,5 & -7.779\,903\,105\,96 &-1.801\,631\,553 &-12.049\,914\,16 & 10.010\,935\,1\\
 12       &-7.478\,060\,323\,889\,7 & -7.779\,903\,104\,98 &-1.801\,631\,587 &-12.049\,913\,45 & 10.010\,940\,4\\
 $\infty$ &-7.478\,060\,323\,906(8) & -7.779\,903\,104\,4 (7) &-1.801\,631\,62 (4) &-12.049\,913\,0 (4) &10.010\,945(4)\\
\hline \multicolumn{6}{c}{Li $3^1S_{1/2}$} \\
 10       &-7.354\,098\,421\,004\,0 & -7.646\,138\,262\,557 &-1.677\,971\,728\,8 &-11.871\,192\,0 &10.014\,596, \\
 11       &-7.354\,098\,421\,302\,1 & -7.646\,138\,262\,527 &-1.677\,971\,758\,0 &-11.871\,177\,9 &10.014\,649 \\
 12       &-7.354\,098\,421\,379\,9 & -7.646\,138\,262\,612 &-1.677\,971\,775\,7 &-11.871\,171\,5 &10.014\,687 \\
 $\infty$ &-7.354\,098\,421\,426 (19) & -7.646\,138\,262\,65 (3) &-1.677\,971\,789 (11) &-11.871\,168(8) &10.014\,72(4)\\
\hline \multicolumn{6}{c}{Li $2^1P_{1/2}$} \\
 10       &-7.410\,156\,532\,150\,2 & -7.656\,895\,306\,5  &-1.806\,107 &-11.801\,371\,0 & 9.685\,503\\
 11       &-7.410\,156\,532\,586\,0 & -7.656\,895\,217\,6  &-1.806\,275  &-11.801\,365\,9 & 9.685\,230\\
 12       &-7.410\,156\,532\,628\,6 & -7.656\,895\,191\,1  &-1.806\,407 &-11.801\,363\,5 & 9.685\,356\\
 $\infty$ &-7.410\,156\,532\,665(14)& -7.656\,895\,176(9)  &-1.806\,51(8) &-11.801\,362(2) & 9.685\,43(8)\\
\hline \multicolumn{6}{c}{Be$^+$ $2^1S_{1/2}$} \\
 10       &-14.324\,763\,176\,616\,3 &-14.777\,682\,315\,84  &-3.634\,056\,961\,9 &-43.688\,039\,3 & 26.932\,300\\
 11       &-14.324\,763\,176\,725\,0 &-14.777\,682\,314\,48  &-3.634\,057\,065\,9 &-43.688\,034\,4 & 26.932\,382\\
 12       &-14.324\,763\,176\,763\,5 &-14.777\,682\,313\,66  &-3.634\,057\,095\,3 &-43.688\,026\,3 & 26.932\,370\\
 $\infty$ &-14.324\,763\,176\,784 (11) &-14.777\,682\,313\,0 (5)&-3.634\,057\,110(11) &-43.688\,023(4) &26.932\,37(2)\\
\hline \multicolumn{6}{c}{Be$^+$ $3^1S_{1/2}$} \\
 10       &-13.922\,789\,268\,385\,7 & -14.351\,840\,883\,53 &-3.294\,071\,80 &-42.344\,339\,8 & 27.308\,080\,8\\
 11       &-13.922\,789\,268\,530\,7 & -14.351\,840\,888\,66 &-3.294\,073\,32 &-42.344\,323\,8 & 27.308\,224\,8\\
 12       &-13.922\,789\,268\,554\,2 & -14.351\,840\,890\,68 &-3.294\,073\,94 &-42.344\,318\,4 & 27.308\,210\,6\\
 $\infty$ &-13.922\,789\,268\,570(10) & -14.351\,840\,891\,8 (8) &-3.294\,074\,4(4) &-42.344\,317(3) & 27.308\,21(2)\\
\hline \multicolumn{6}{c}{Be$^+$ $2^1P_{1/2}$} \\
 10       &-14.179\,333\,292\,319\,7 &-14.345\,507\,657 &-3.728\,487 &-42.123\,997\,4 & 24.352\,898\\
 11       &-14.179\,333\,293\,202\,9 &-14.345\,507\,560 &-3.728\,679 &-42.123\,985\,4 & 24.352\,585\\
 12       &-14.179\,333\,293\,342\,7 &-14.345\,507\,534 &-3.728\,739 &-42.123\,978\,2 & 24.352\,564\\
 $\infty$ &-14.179\,333\,293\,42(3) &-14.345\,507\,52(3)&-3.728\,78(4)&-42.123\,976(6) & 24.352\,55(4)\\

\end{tabular}
\end{ruledtabular}
\end{table}
\end{widetext}

QED corrections include two new terms, $1/r^3$ and the Bethe logarithm,
see Eqs. (\ref{08}) and (\ref{08p}).
Matrix elements of the $1/r^3$ term involve Hylleraas integrals
with $n_i=-2$. Their calculation using recursion relations was presented in
Ref. \cite{rec3}, and numerical results are presented in Table \ref{table2}. 
We note a slow numerical convergence of $1/r^3$ terms and its mass polarization correction.
Bethe logarithms
are far the most difficult in accurate numerical evaluation and in this
work we use the results obtained by Yan {\em et al.} in \cite{dy_bethe2}. 
We note a weak dependence of Bethe logarithms on a state, 
all close to the 1S hydrogenic value.
\begin{widetext}
\squeezetable
\begin{table}[!hbt]
\renewcommand{\arraystretch}{1.0}
\caption{Relativistic and QED operators in Li and Be$^+$.
         Implicit sum over $a$ and sum over $a\neq b$ pairs are assumed.
         Bethe logarithms are that from Yan {\em et al.} \cite{dy_bethe2} }
\label{table2}
\begin{ruledtabular}
\begin{tabular}{lrrrrrr}
operator  & \cent{{\rm Li}\;2S_{1/2}} & \cent{{\rm Li}\;3S_{1/2}}   & \cent{{\rm Li}\;2P_{1/2}}
          & \cent{{\rm Be}^+\;2S_{1/2}} & \cent{{\rm Be}^+\;3S_{1/2}}   & \cent{{\rm Be}^+\;2P_{1/2}} \\
\hline
 $p_a p_b$
                 &0.301\,842\,780\,3(6)&0.292\,039\,841\,2(2)&0.246\,738\,644(9)
                 &0.452\,919\,136\,1(7)&0.429\,051\,623\,1(7)&0.166\,174\,23(2)\\
 $p_a p_b/(E-H)' p_c p_d$
                 &-1.499\,788\,83(4)&-1.385\,931\,96(2)&-1.559\,77(7)
                 &-3.181\,137\,97(3)&-2.865\,022\,8(3)&-3.562\,60(5)\\
 $p_a^4$
                 &628.449\,022(4)&622.859\,40(2)&620.044\,977(8)
                 &2146.520\,76(5)&2108.368\,02(9)&2088.652\,95(4)\\
 $[p_a^4]_{\rm mp}$
                 &41.176\,62(5)&40.628\,83(7)&40.628\,87(6)
                 &115.923\,1(2)&112.589\,8(3)&154.322\,2(4)\\
 $p_a^i\Bigl(\frac{\delta^{ij}}{r_{ab}}+\frac{r_{ab}^i\,r_{ab}^j}{r_{ab}^3}\Bigr)\,p_b^j$
                 &0.871\,195\,809 (9)&0.859\,817\,43(18)&0.792\,851\,59(4)
                 &1.819\,804\,86(12)&1.774\,364\,56(14)&1.226\,817\,28(11)\\
 $\Bigl[p_a^i\Bigl(\frac{\delta^{ij}}{r_{ab}}+\frac{r_{ab}^i r_{ab}^j}{r_{ab}^3}\Bigr)p_b^j\Bigr]_{\rm mp}$
                 &6.154\,303\,2(3)&5.844\,828\,2(5)&6.002\,973(11)
                 &17.854\,151(2)&16.414\,367(3)&17.777\,145(11)\\
 $\delta^3 (r_a)$
                 &13.842\,610\,787(6)&13.736\,502\,84(7)&13.676\,197\,06(7)
                 &35.105\,055\,72(8)&34.577\,877\,6(1)&34.245\,820\,96(5)\\
 $[\delta^3 (r_a)]_{\rm mp}$
                 &0.484\,589\,3(6)&0.487\,894\,7(6)&0.647\,914(5)
                 &0.927\,607\,3(4)&0.932\,129 (2)&1.831\,450(7)\\
 $\delta^3 (r_{ab})$
                 &0.544\,324\,632\,0(7)&0.536\,168\,418\,9(4)&0.532\,274\,098\,9(3)
                 &1.580\,538\,588(3)&1.537\,328\,373(1)&1.518\,990\,086(6)\\
 $[\delta^3 (r_{ab})]_{\rm mp}$
                 &-0.082\,009\,7(4)&-0.078\,406(2)&-0.066\,725(2)
                 &-0.431\,732(2)&-0.409\,936\,9(9)&-0.351\,752(3)\\
 $p_N^i\Bigl(\frac{\delta^{ij}}{r_{a}}+\frac{r_{a}^i\,r_{a}^j}{r_{a}^3}\Bigr)\,p_a^j$
                 &87.276\,740\,9(3)&86.289\,222\,5(7)&85.743\,261\,2(7)
                 &222.628\,511\,9(4)&217.799\,579\,5(7)&214.562\,695\,2(8)\\
 $1/r_{ab}^3$
                 &0.273\,413(5)&0.198\,05(8)&0.289\,57(5)
                 &-7.514\,6(1)&-7.579\,45(7)&-6.794\,2(2)\\
 $[1/r_{ab}^3]_{\rm mp}$
                 &-1.627\,46(5)&-1.645\,4(2)&-1.892(5)
                 &-7.514\,6(1)&-7.579\,45(7)&-6.794\,2(2)\\
 $1/r_{a}^3$
                 &-308.314\,23(6)&-305.939\,0(2)&-304.517\,7(2)
                 &-910.919\,4(1)&-897.084\,8(3)&-887.841\,7(3)\\
 $\ln(k_0)-2\ln Z$ 
                 &2.981\,06(1)&2.982\,36(6)&2.982\,57(6)
                 &2.979\,26(2)&2.981\,62(1)&2.982\,27(6)\\
 $[\ln(k_0)]_{\rm mp}$ 
                 &-0.113\,05(5)&-0.110\,5(3)&-0.111\,2(5)
                 &-0.125\,58(4)&-0.117\,1(1)&-0.121\,7(6)\\
 $\vec r_a/r_a^3\times\vec p_a\cdot\vec\sigma_a$
                 &&&-0.125\,946\,352(50)&&&-0.969\,131\,7(11)\\
 $[\vec r_a/r_a^3\times\vec p_a\cdot\vec\sigma_a]_{\rm mp}$
                 &&&0.376\,388(3) &&& 3.043\,394(15)\\
 $\vec r_{ab}/r_{ab}^3\times\vec p_a\cdot\vec\sigma_a$
                 &&&-0.224\,640\,70(5) &&&-1.659\,492\,5(3)\\
 $[\vec r_{ab}/r_{ab}^3\times\vec p_a\cdot\vec\sigma_a]_{\rm mp}$
                 &&&0.570\,585(4) &&& 4.532\,68(9) \\
 $\vec r_{ab}/r_{ab}^3\times\vec p_b\cdot\vec\sigma_a$
                 &&&0.038\,473\,60(6)&&& 0.360\,851\,6(2) \\
 $[\vec r_{ab}/r_{ab}^3\times\vec p_b\cdot\vec\sigma_a]_{\rm mp}$
                 &&&-0.213\,52(3) &&& -1.549\,82(10) \\
 $\vec r_a/r_a^3\times\vec p_N\cdot\vec\sigma_a$
                 &&&0.022\,524\,93(9) &&& 0.339\,008\,2(2)\\
\end{tabular}
\end{ruledtabular}
\end{table}
\end{widetext}

\section{Transition and ionization energies}
Numerical values of all coefficients for relevant
transition energies are presented for lithium and beryllium in the Table
\ref{table3}. ${\cal E}^{(5,0)}$ does not
include there the nuclear polarizability contribution. It is a small
correction, which results from modification of the nucleus by atomic electrons.
The effect of nuclear polarizability in $^{11}$Li was estimated in \cite{mp_li_iso},
and $^{11}$Be$^+$ is obtained here. While it does not affect the 
absolute transition frequencies much, it is significant for
the isotope shift and the corresponding determination of the charge radii
difference, see the next section.

Obtained results for the energy expansion coefficients
are in general in good agreement with the former
calculation of Yan and Drake in \cite{dy_li_iso1, dy_li_iso2, dy_bethe1},
with corrections and improvements in \cite{dy_bethe2}.
We have not yet confirmed their results for the Bethe logarithms \cite{dy_bethe2}
and use them in our coefficients ${\cal E}^{(5,0)}$ and ${\cal E}^{(5,1)}$.
We note that the present numerical precision of expansion coefficients
is high enough, that the leading uncertainty of transition frequencies
comes from ${\cal E}^{(6,0)}$, more precisely from the rough estimate of
the relativistic (nonradiative) $m\,\alpha^6$ correction,
which is about 10\% of the leading radiative correction.
\begin{widetext}
\begin{table}[!hbt]
\renewcommand{\arraystretch}{1.0}
\caption{Expansion of energy in powers of $\alpha$ and $\eta$ in Li. The last
  column presents values in units cm$^{-1}$ for the ground state ionization
  energy for $^7$Li and $^9$Be$^+$ with atomic masses from Table \ref{table4}. 
  ${\cal E}^{(4,0)}_{\rm fs}$ is the finite size correction
  with $r_{\rm ch}(^7{\rm Li}) = 2.39(3)$ fm \cite{liradius} 
  and $r_{\rm ch}(^9{\rm Be}) =  2.519(12)$ fm \cite{beradius}.}
\label{table3}
\begin{ruledtabular}
\begin{tabular}{cw{3.15}w{2.15}w{2.17}w{6.14} }
energy          & \cent{3S_{1/2}-2S_{1/2}}   &
\cent{2P_{1/2}-2S_{1/2}} & \cent{\rm{I.P}. \; 2S_{1/2}} & \cent{\rm{I.P}. \; 2S_{1/2}\;[{\rm cm}^{-1}] } \\
\hline \multicolumn{5}{c}{Li}\\
 ${\cal E}^{(2,0)}$  & 0.123\,961\,902\,48(2)  & 0.067\,903\,791\,24(2)& 0.198\,146\,911\,238(12) &  43\,488.220\,301(6) \\
 ${\cal E}^{(2,1)}$  & 0.133\,764\,841\,8(7)   & 0.123\,007\,928(10)   & 0.211\,013\,905\,1(6)    & -3.621\,708 \\
 ${\cal E}^{(2,2)}$  & 0.123\,659\,827(11)     &-0.004\,88(7)          & 0.235\,270\,010(19)      &  0.000\,316 \\
 ${\cal E}^{(4,0)}$  & 0.178\,745(2)           & 0.248\,551\,4(10)     & 0.240\,540\,1(4)         &  2.811\,269(5)\\
 ${\cal E}^{(4,0)}_{\rm fs}$  & -0.666\,695 \langle r^2_{\rm ch}\rangle
                            & -1.045\,608 \langle r^2_{\rm ch}\rangle
                            & -0.870\,799 \langle r^2_{\rm ch}\rangle
                            & -0.000\,390  \\
 ${\cal E}^{(4,1)}$  & 0.003\,78(4)            &-0.325\,44(11)         & 0.013\,50(3)             & -0.000\,012\\
 ${\cal E}^{(5,0)}$  &-2.193(3)                &-3.478(3)              &-2.860\,8(6)              & -0.243\,99(5)    \\
 ${\cal E}^{(5,1)}$  &-1.46(2)                 & 1.30(3)               &-1.889 (3)                &  0.000\,013 \\
 ${\cal E}^{(6,0)}$  &-9.2(9)                  &-14.4(1.4)             &-12.0(1.2)                & -0.007\,5(7)    \\
 ${\cal E}^{(6,1)}$  & -38.(10)                &-46.(12)               &-49.(12)                  &  0.000\,002    \\
 ${\cal E}^{(7,0)}$  & 115.(29)                & 215.(54)              & 147.(37)                 &  0.000\,67(17)\\
\hline \multicolumn{5}{c}{Be$^+$}\\
 ${\cal E}^{(2,0)}$  & 0.401\,973\,908\,21(2)  & 0.145\,429\,883\,35(5) &0.669\,196\,938\,370(19)& 146\,871.751\,363(4)\\
 ${\cal E}^{(2,1)}$  & 0.425\,841\,421\,2(13)  & 0.432\,174\,796(11)    &0.701\,595\,771\,3(4)   & -9.374\,767        \\
 ${\cal E}^{(2,2)}$  & 0.339\,982\,7(3)        & -0.094\,72(3)          &0.721\,945\,175(15)     &  0.000\,587 \\
 ${\cal E}^{(4,0)}$  & 1.343\,705(3)           & 1.564\,047(3)          &1.892\,557(3)           &  22.118\,92(4)\\
 ${\cal E}^{(4,0)}_{\rm fs}$  &-4.416\,477(1)\langle r^2_{\rm ch}\rangle
                            &-7.198\,308(1)\langle r^2_{\rm ch}\rangle
                            &-5.937\,280(1)\langle r^2_{\rm ch}\rangle
                            & -0.002\,955  \\
 ${\cal E}^{(4,1)}$  & 0.375\,84(2)            &-2.579\,82(5)           &0.620\,19(3)            & -0.000\,441  \\
 ${\cal E}^{(5,0)}$  &-13.168(2)               &-21.864(11)             &-17.742(4)              & -1.513\,1(3) \\
 ${\cal E}^{(5,1)}$  & 3.19(2)                 & 30.92(13)              & 28.75(8)               & -0.000\,149 \\
 ${\cal E}^{(6,0)}$  &-81.(8)                  &-132.(13)               &-109.(11)               & -0.068(7)   \\
 ${\cal E}^{(6,1)}$  &-361.(90)                &-442.(111)              &-477.(119)              &  0.000\,018(5) \\
 ${\cal E}^{(7,0)}$  & 1036.(259)              & 1787.(447)             &1360.(340)              &  0.006\,2(15)   \\
\end{tabular}
\end{ruledtabular}
\end{table}
\end{widetext}
Using this expansion coefficients in Table \ref{table3}, 
and atomic masses from Table \ref{table4}, 
\begin{table}[!hbt]
\renewcommand{\arraystretch}{1.0}
\caption{Atomic masses of Lithium and Beryllium isotopes and the atomic
  binding energy.}
\label{table4}
\begin{ruledtabular}
\begin{tabular}{llllll}
Li isotope      & \centt{mass [u]} & Ref. & Be isotope      & \centt{mass [u]}& Ref. \\
\hline
$^6$Li&6.015122794(16) &\cite{li_mass}   & $^7$Be      &7.016\,929\,83(11) &\cite{nu_mass} \\
$^7$Li&7.0160034256(45) &\cite{7limass}  & $^9$Be      & 9.012\,182\,20(43) &\cite{nu_mass}   \\
$^8$Li&8.02248624(12) &\cite{li_mass}  & $^{10}$Be   &10.013\,533\,82(43) &\cite{nu_mass} \\
$^9$Li&9.02679020(21) &\cite{li_mass}   & $^{11}$Be   &11.021\,661\,55(63) &\cite{11bemass}   \\
$^{11}$Li&11.04372361(69) &\cite{li_mass} & $^{14}$Be   &14.042\,890(140) &\cite{nu_mass}    \\
$E_{\rm Li}$&-7.281\; {\rm au}   &  & $E_{\rm Be}$ & -14.669\; {\rm au}
\end{tabular}
\end{ruledtabular}
\end{table}
one obtains transition frequencies which are compared in Table \ref{table5} to
the previous calculations of Yan {\em et al.} \cite{dy_li_iso1,
dy_li_iso2, dy_bethe2} and to the experimental results. Small
differences with results of Drake and Yan are due to the
better numerical accuracy of our results and the inclusion of the
finite nuclear size correction in transition and ionization energies
for Be$^+$. In comparison to
experimental values we observe an agreement for both the lithium atom
and beryllium ion, with one exception. Namely, theoretical
ionization energy is larger by $0.06$ cm$^{-1}$ than the
experimental value, and this discrepancy was already pointed out in
\cite{dy_bethe2}. The relatively lower accuracy of theoretical
results for the beryllium ion comes from the neglect of the
nonradiative $m\,\alpha^6$ correction, which significantly grows
with $Z$. The direct calculation of this correction for the three
electron system is a challenge. A simpler approach would rely on
matching the high Z results for lithiumlike systems \cite{yerokhin1}
with the low Z results obtained by the expansion in $\alpha$. We note that
in spite of the relatively large uncertainties coming from ${\cal E}^{(6,0)}$ the
obtained results for transition frequencies are the most accurate so
far, and no other approach allows one for the systematic calculation of
all corrections in the low $Z$ atomic systems.
\begin{widetext}
\begin{table}[!hbt]
\renewcommand{\arraystretch}{1.0}
\caption{Comparison of our theoretical predictions with the previous
  theoretical and experimental values in units cm$^{-1}$. 
  References: a-\cite{dy_bethe2}, b-\cite{bushaw1}, c-\cite{bushaw2},
  d-\cite{sansonetti}, e-\cite{ralchenko}}
\label{table5}
\begin{ruledtabular}
\begin{tabular}{llll}
 & \cent{\rm{I.P}. \;  2S_{1/2}}        & \cent{3S_{1/2}-2S_{1/2}}   & \cent{2P_{1/2}-2S_{1/2}}\\
\hline
 $^7{\rm Li}$ (this work) & 43\,487.159\,0(8)      & 27\,206.093\,7(6)       & 14\,903.648\,4(10)        \\
 $^7{\rm Li}$ (the)       & 43\,487.158\,3(10)$^a$ & 27\,206.093\,0(10)$^a$  & 14\,903.647\,9(10)$^a$  \\
 $^7{\rm Li}$ (exp)       & 43\,487.159\,40(18)$^b$& 27\,206.094\,20(10)$^c$ & 14\,903.648\,130(14)$^d$\\ \hline
 $^9{\rm Be}^+$(this work)& 146\,882.918(7)        & 88\,231.919(5)          & 31\,928.734(8)    \\
 $^9{\rm Be}^+$ (the)     & 146\,882.923(5)$^a$    & 88\,231.920(6)$^a$      & 31\,928.738(5)$^a$        \\
 $^9{\rm Be}^+$ (exp)     & 146\,882.86$^e$        & 88\,231.915$^e$         & 31\,928.744$^e$       \\
\end{tabular}
\end{ruledtabular}
\end{table}
\end{widetext}

\section{Isotope shift determination of nuclear charge radii}
The isotope shift in atomic transitions come mainly from different masses of
nuclei. Much smaller effect of order $\sim 10^{-6}$ is due to different
nuclear charge radii. Nevertheless, precision of isotope shift measurements and theoretical
predictions is enough to derive charge radii from the comparison of
the experimental and the theoretical data. Moreover, this determination of charge radii
is far more accurate than the one obtained from electron scattering off nuclei
and can be applied even to unstable nuclei such as $^{11}$Li and $^{11}$Be.
From expansion coefficients for Li and Be$^+$ in the Table \ref{table3},
one obtains contributions to the isotope shift. The example for
$^{11}$Li-$^{7}$Li and $^{11}$Be$^+$-$^{9}$Be$^+$ is presented in Table \ref{table7}.
We observe that the leading nonrelativistic contribution gives at least 99.9\%
of the total isotope shift. The relativistic recoil corrections
are small but still important, while the theoretical uncertainty is
dominated by rough estimation of ${\cal E}^{(6,1)}$. The nuclear polarizability
correction is significant for both $^{11}$Li and $^{11}$Be$^+$,
and its value is presented in Table \ref{table7}. Result from $^{11}$Li
was obtained in \cite{mp_li_iso}, while for $^{11}$Be$^+$ it is calculated here using
both experimental \cite{palit} and theoretical data \cite{typel} for the so called $B(E1)$
function which is the reduced line strength for the nuclear $E1$ excitation,
\begin{equation}
|\langle\phi_N|\vec d|E\rangle|^2 = \frac{4\,\pi}{3}\,\frac{dB(E1)}{dE}, \label{p04}
\end{equation}
in units $e^2$ fm$^2$ MeV$^{-1}$. The kets $|\phi_N\rangle$ and $| E \rangle$ denote
the ground state of the nucleus and the excited state with excitation
energy $E$, respectively.
The relevant formula relating electric dipole nuclear 
transition moment with the shift of
atomic energy levels is \cite{mp_li_iso},
\begin{equation}
\nu_{\rm pol} = -m\,\alpha^4\,\Bigl\langle\sum_a\delta^3(r_a)\Bigr\rangle\;
(m^3\,\tilde\alpha_{\rm pol}), \label{p02}
\end{equation}
where $\tilde\alpha_{\rm pol}$  is a {\em weighted} electric polarizability of the nucleus
and is given by the following double integral
\begin{eqnarray}
\tilde\alpha_{\rm pol} &=& \frac{16\,\alpha}{3}\,\int dE\,
\frac{1}{e^2}\,|\langle\phi_N|\vec d|E\rangle|^2\,\int_0^\infty\,\frac{d w}{w}\,
\frac{E}{E^2+w^2}
\nonumber \\ &&\hspace*{-5ex}
\times\frac{1}{(\kappa+\kappa^\star)}\,\biggl[1+
\frac{1}{(\kappa+1)(\kappa^\star+1)}\,\biggl(\frac{1}{\kappa+1}+\frac{1}{\kappa^\star+1}
\biggr)\biggr],\label{p03}
\end{eqnarray}
where $ \kappa = \sqrt{1+2\,i\,m/w}$. The first integral over the nuclear
excitation spectrum may involve a sum over discreet levels, as it is
the case of $^{11}$Be. This nucleus has an excited state with $E=0.320$ MeV
and $B(E1) = 0.116\;e^2$ fm$^2$ and a continuum spectrum starts at $E_T =
0.504$ MeV. The result for $\tilde\alpha_{\rm pol}$ using the experimental
\cite{palit} or the theoretical data \cite{typel} is 
\begin{equation}
\tilde\alpha_{\rm pol} = 39.7(40)\;{\rm fm}^3 =  6.90(69)\;10^{-7}\;m^{-3},
\end{equation}
and this value is used to obtain the shift of energy levels $\nu_{\rm pol}$ in
Eq. (\ref{02}) for Be$^+$ in Table \ref{table7}.
\begin{table}[!hbt]
\renewcommand{\arraystretch}{1.0}
\caption{Contributions to the $^{11}$Li - $^{7}$Li isotope shift of $3S_{1/2}-2S_{1/2}$ transition,
  and to the $^{11}$Be$^+$ - $^{9}$Be$^+$ shift of $2P_{1/2}-2S_{1/2}$ transition, with excluding
  the finite size correction. The second uncertainty of $\Delta \nu_{\rm the}$
  is due to the atomic mass.}
\label{table7}
\begin{ruledtabular}
\begin{tabular}{lw{6.12}w{6.11}}
correction      & \centt{Li($3S_{1/2}-2S_{1/2}$) [MHz]} & \centt{Be$^+$($2P_{1/2}-2S_{1/2}$) [MHz]} \\
\hline
$\Delta \nu^{(2,1)}$           & 25\,104.520\,2(1)&  31\,568.577\,3(8) \\
$\Delta \nu^{(2,2)}$           & -2.967\,9      &    0.765\,7(2) \\
$\Delta \nu^{(4,1)}$           &  0.037\,8(4)      &  -10.035\,0(2) \\
$\Delta \nu^{(5,1)}$           & -0.106\,4(15)  &    0.877\,7(36)\\
$\Delta \nu^{(6,1)}$           & -0.020(5)      &   -0.092(23)\\
$\Delta \nu_{\rm pol}$         &  0.039(4)      &    0.208(21)\\[1ex]
$\Delta \nu_{\rm the}$         & 25~101.502\,8 (64)(27)& 31\,560.302(31)(12)\\
$\Delta \nu_{\rm the}$ \cite{dy_bethe2}        & 25~101.470(22)& 31\,560.01(6)\\
\end{tabular}
\end{ruledtabular}
\end{table}

From the difference between experimental and theoretical isotope shift
one determines nuclear charge radii by using
\begin{equation}
\Delta\nu_{\rm exp}-\Delta\nu_{\rm the}
= C_{AB}\,(r_{{\rm ch}\,A}^2-r_{{\rm ch}\,B}^2),
\end{equation}
with constant $C$ obtained from Eq. (\ref{06}) with including
logarithmic relativistic corrections to the wave function at the origin
\begin{equation}
C = \frac{2 \pi}{3} Z \alpha^4\,\Bigl\langle\sum_a\delta^{(3)}(r_a)\Bigr\rangle\,
\bigl[1-(Z\,\alpha)^2\,\ln(Z\,\alpha\,m\,r_{\rm ch})\bigr].
\end{equation}
Using isotope shifts as measured for Li in \cite{lit_iso1, lit_iso2} and Be$^+$ in \cite{be_iso},
we obtain nuclear charge radii for corresponding isotopes in Tables
\ref{table8} and \ref{table9}.
Results for Li are slightly more accurate than our previous determination in
\cite{mp_li_iso} due to the more accurate $\nu_{\rm the}$ and nuclear masses.
Our results for Be isotopes, agree with the recent
determination presented in \cite{be_iso}. The uncertainty of our $\Delta
\nu_{\rm the}$ comes mainly from $25\%$ of $\Delta \nu^{(6,1)}$
and $10\%$ of $\Delta\nu_{\rm pol}$. Nonnegligible are numerical uncertainties
of Bethe logarithms and their mass polarization corrections.
The uncertainty of $C$ coefficients comes from the estimation of 
relativistic correction to the wave function at origin, which is 
about $25\%$ of the logarithmic part. Nevertheless, uncertainties in 
$\delta r^2_{\rm ch}$ come mainly from the experimental value for the isotope shift,
and the uncertainty of the final $r_{\rm ch}$ comes mostly from
the charge radius of the reference nucleus. As we have already mentioned,
the direct determination of the charge radius from the absolute transition
frequency is at present not possible for lithiumlike systems due to
insufficient precision of theoretical predictions.
\begin{widetext}
\begin{table}[!htb]
\begin{minipage}{16.0cm}
\renewcommand{\arraystretch}{1.3}
\caption{Summary of isotope shift determination of Li charge radii from
  $3S_{1/2}-2S_{1/2}$ transition with respect to $^7$Li, 
$r(^7{\rm Li}) =  2.39(3)$ fm \cite{liradius},  the first uncertainty of $\nu_{\rm the}$ comes
from unknown higher order terms, the second uncertainty is due to the atomic mass.}
\label{table8}
\begin{ruledtabular}
\begin{tabular}{rw{6.8}w{7.12}w{2.8}w{2.7}w{1.7}}
      isotope & \centt{$\nu_{\rm exp}$[MHz] \cite{lit_iso2}}  &
      \centt{$\nu_{\rm the}$[MHz]} & \centt{$C$ [MHz fm$^{-2}$]} &
      \centt{$\delta r^2_{\rm ch}$[fm$^2$]}& \centt{$r_{\rm ch}$[fm]} \\ \hline
$^6$Li   &-11~453.983(20)  & -11~452.820\,5 (23)(2) & -1.571\,9(16) &  0.740(13)  &  2.540(28)\\
$^8$Li   &  8~635.782(44)  &   8~634.981\,2 (17)(9) & -1.572\,0(16) & -0.509(28)  &  2.281(32)\\
$^9$Li   & 15~333.272(39)  &  15~331.799\,5 (31)(12)& -1.572\,1(16) & -0.937(25)  &  2.185(33)\\
$^{11}$Li& 25~101.226(125) &  25~101.502\,8 (64)(27)& -1.576\,8(17) &  0.176(79)  &  2.426(34)\\
\end{tabular}
\end{ruledtabular}
\end{minipage}
\end{table}
\end{widetext}

\begin{widetext}
\begin{table}[!htb]
\begin{minipage}{16.0cm}
\renewcommand{\arraystretch}{1.3}
\caption{Summary of isotope shift determination of Be$^+$ charge radii from
  $2P_{1/2}-2S_{1/2}$ transition with respect to $^9$Be$^+$,
  $r(^9{\rm Be}) =  2.519(12)$ fm \cite{beradius}, the first uncertainty of $\nu_{\rm the}$
  comes from unknown higher order terms, the second uncertainty is due to the atomic mass.}
\label{table9}
\begin{ruledtabular}
\begin{tabular}{rw{6.6}w{6.11}w{2.7}w{2.7}w{1.7}}
      isotope & \centt{$\nu_{\rm exp}$[MHz] \cite{be_iso}}  & \centt{$\nu_{\rm the}$[MHz]}
             & \centt{$C$ [MHz fm$^{-2}$]} & \centt{$\delta r^2_{\rm ch}$[fm$^2$]}
             & \centt{$r_{\rm ch}$[fm]} \\ \hline
$^7{\rm Be}^+$   &  -49~236.81(88) & -49\,225.736(35)(9)  & -17.021(31) &  0.651(47)  & 2.645(14) \\
$^{10}{\rm Be}^+$&   17~323.8(13) &  17\,310.437(13)(11) & -17.027(31) & -0.785(76)  & 2.358(21)\\
$^{11}{\rm Be}^+$&   31~564.96(93) &  31\,560.302(31)(12) & -17.020(31) & -0.274(55)  & 2.464(16) \\
\end{tabular}
\end{ruledtabular}
\end{minipage}
\end{table}
\end{widetext}

\section{Summary}
We have calculated nonrelativistic, relativistic, and QED contributions
to low-lying energy levels of Li and Be$^+$ ions, including finite nuclear
mass corrections. The computational method uses the Hylleraas basis set with
the analytic integration technique. The obtained results are the most precise
to date, with the accuracy limited mainly by higher order relativistic
$m\,\alpha^6$ corrections. Using the experimental results for the
isotope shift of $3S_{1/2} - 2S_{1/2}$ transition in Li \cite{lit_iso1, lit_iso2} and
$2P_{1/2} - 2S_{1/2}$ transition in Be$^+$ \cite{be_iso}, we obtain improved
charge radii for Li and Be isotopes. We note the significance
of the nuclear polarizability effect in $^{11}$Li and $^{11}$Be$^+$
and relativistic correction to the wave function at origin
for the determination of charge radii.

The presented computational method is limited by unknown higher order
relativistic and QED corrections, which become more significant for
heavier nuclei. One possible solution for the charge radii determination
for heavier nuclei is the spectrosopy of the four-electron ion, for which we think,
accurate calculations can be performed with the help of a Gaussian basis set
with linear terms \cite{lgaus1,lgaus2}.
Apart from nuclear charge radii, precise atomic spectroscopy may bring
information about the magnetic moment distribution within nuclei.
Indeed a measurement of the hyperfine splitting in $^{11}$Be$^+$ \cite{nak}
may give the size of neutron halo, which cannot be probed by other means.
However, the interpretation of the shift of the hyperfine splitting in terms of
the Bohr-Weiskopf effect is not obvious, due to possible large nuclear
polarizability effects \cite{vecpol}.

The significant advantage of the presented computational approach
with Hyllerras functions is the ability to calculate
higher order relativistic \cite{lihfs} and QED corrections, 
although such a calculation is not simple. 
We aim to obtain $O(\alpha^2)$ and  $O(\alpha^3)$
corrections to the hyperfine splitting in order to investigate nuclear
structure correction with halo nuclei,
and also to verify the accuracy of simplified approaches such as 
the relativistic configuration interaction or the multiconfiguration Dirac-Fock
method.

\section*{Acknowledgments}
We wish to thank Krzysztof Rusek for helpful information on the Be nuclei,
Vladimir Yerokhin for valuable coments,
Stephan Typel for sending us numerical data for the $B(E1)$
function in $^{11}$Be, and Willfried N\"ortersh\"auser
for sending us his work prior to publication. 
Authors acknowledge support by NIST
Precision Measurement Grant No. PMG 60NANB7D6153.


\begin{thebibliography}{99}
\bibitem{lindgren} I. Lindgren, Mol. Phys. {\bf 98}, 1159 (2000).
\bibitem{shab} V. Shabaev, Phys. Rep. {\bf 356}, 119 (2002).
\bibitem{h6hel} K. Pachucki, Phys. Rev. A {\bf 74}, 022512 (2006),
                {\bf 74}, 062510, (2006), {\bf 76} 059906(E) (2007).
\bibitem{drake_fs} G.W.F. Drake, Can. J. Phys. {\bf 80}, 1195 (2002).
\bibitem{krp_fs} K. Pachucki, Phys. Rev. Lett. {\bf 97}, 013002 (2006).
\bibitem{dy_bethe1}  Z.-C. Yan and G. W. F. Drake, Phys. Rev. Lett. {\bf 91}, 113004 (2003).
\bibitem{dy_bethe2}  Z.-C. Yan, W. N\"ortersh\"auser, and G. W. F. Drake,
                     Phys. Rev. Lett. {\bf 100}, 243002 (2008).
\bibitem{beqed} K. Pachucki and J. Komasa, Phys. Rev. Lett. {\bf 92}, 213001 (2004).
\bibitem{hd_iso} A. Huber {\em et al.}, Phys. Rev. Lett. {\bf 80}, 468 (1998).
\bibitem{he6_exp} L.-B. Wang {\em et al.}, Phys. Rev. Lett. {\bf 93}, 142501 (2004).
\bibitem{he8_exp} P. Mueller {\em et al.}, Phys. Rev. Lett. {\bf 99}, 252501 (2007).
\bibitem{lit_iso1} G. Ewald {\em et al.}, Phys. Rev. Lett. {\bf 93}, 113002 (2004).
\bibitem{lit_iso2} R. S\'anchez {\em et al.}, Phys. Rev. Lett. {\bf 96}, 033002 (2006).
\bibitem{dy_li_iso1} Z-C. Yan and G. W. F. Drake, Phys. Rev. A {\bf 61}, 022504 (2000).
\bibitem{dy_li_iso2} Z-C. Yan and G. W. F. Drake, Phys. Rev. A {\bf 66}, 042504 (2002).
\bibitem{mp_li_iso} M. Puchalski, A.M. Moro, and K. Pachucki,
                    Phys. Rev. Lett {\bf 97}, 133001 (2006).
\bibitem{yerokhin2} V. A. Yerokhin, P. Indelicato, and V. M. Shabaev,
                    Phys. Rev. A {\bf 71}, 040101(R) (2005). 
\bibitem{king_lit} F.W. King, J. Mol. Struct. {\bf 400}, 7 (1997).
\bibitem{yan_lit} Z.-C. Yan and G.W.F. Drake, Phys. Rev. A {\bf 52}, 3711 (1995).
\bibitem{rec1} K. Pachucki, M. Puchalski and E. Remiddi,
                    Phys. Rev. A {\bf 70}, 032502 (2004).
\bibitem{rec2} K. Pachucki and M. Puchalski, Phys. Rev. A {\bf 71}, 032514 (2005).
\bibitem{li_ground} M. Puchalski and K. Pachucki, Phys. Rev. A {\bf 73}, 022503 (2006).
\bibitem{rec3} K. Pachucki and M. Puchalski, Phys. Rev. A {\bf 77}, 032511 (2008).
\bibitem{sansonetti} C.J. Sansonetti, B. Richou, R. Engleman, Jr.,
                     and L.J. Radziemski, Phys. Rev. A {\bf 52}, 2682 (1995). 
\bibitem{bushaw1} B.A. Bushaw, W. N\"ortersh\"auser, G. Ewald, A. Dax,
                  and G.W.F. Drake, Phys. Rev. Lett. {\bf 91}, 043004 (2003). 
\bibitem{bushaw2} B.A. Bushaw, W. N\"ortersh\"auser,
                  G.W.F. Drake, and H.-J.Kluge, Phys. Rev. A {\bf 75}, 052503 (2007). 
\bibitem{ralchenko} Yu. Ralchenko, A.E. Kramida, J. Reader,
                    and NIST ASD Team, NIST Atomic Spectra Database (version
                    3.1.4) (2008). Available:  http://physics.nist.gov/asd3.
\bibitem{be_iso} W. N\"ortersh\"auser {\em et al.}, submitted to Phys. Rev. Lett. (2008).
\bibitem{darwin} I.B. Khriplovich, A.I. Milstein, and R.A. Sen'kov, J. High Energ. Phys. {\bf 14} 175, (1999).
\bibitem{friar} J.L. Friar, J. Martorell, D.W.L. Sprung, Phys. Rev. A {\bf 56}, 4579 (1997).
\bibitem{bs} H.A.~Bethe and E.E.~Salpeter,
              {\em Quantum Mechanics Of One- And Two-Electron Atoms}, (Plenum, New York, 1977).
\bibitem{drake_book} {\em Handbook of Atomic, Molecular and Optical Physics},
                     edited by G.~W.~F. Drake, (Springer, New York, 2006).
\bibitem{simple} K. Pachucki, J. Phys. B {\bf 31}, 5123 (1998).
\bibitem{palit} R. Palit {\em et al.}, Phys. Rev. C {\bf 68}, 034318 (2003).
\bibitem{typel} S. Typel and G. Baur, Phys. Rev. Lett. {\bf 93}, 142502 (2004).
\bibitem{fw} K. Pachucki, Phys. Rev. A {\bf  71}, 012503 (2005).
\bibitem{eides} M.I. Eides, H. Grotch, and V.A. Shelyuto, Phys. Rep. {\bf 342}, 63 (2001).
\bibitem{korobov} V. Korobov, {\em private communication}.
\bibitem{liradius} C.W. de Jager, H. deVries, and C. deVries, At. Data Nucl. Data
                Tables {\bf 14}, 479 (1974).
\bibitem{beradius} J.A. Jansen, R.Th. Peerdeman, and C. deVries,
                 Nucl. Phys. A {\bf 188}, 337 (1972).
\bibitem{li_mass} M. Smith {\em et al.}, e-print arXiv:0807.1260,
                  Phys. Rev. Lett. (to be published)
\bibitem{nu_mass} G. Audi, A.H. Wapstra, and C. Thibault, Nucl. Phys A {\bf
                   729}, 337 (2003), http://amdc.in2p3.fr /web/masseval.html.
\bibitem{7limass} S. Nagy, {\em et al}., Phys. Rev. Lett. {\bf 96}, 163004 (2006).
\bibitem{11bemass} R. Ringle {\em et al.}, (unpublished).
\bibitem{yerokhin1} V.~A.~Yerokhin, A.~N.~Artemyev, V.~M.~Shabaev,
                   M.~M.~Sysak, O.~M.~Zherebtsov, and G.~Soff,
                   Phys. Rev. Lett. {\bf 85}, 4699 (2000). 
\bibitem{lgaus1} K. Pachucki and J. Komasa, Chem. Phys. Lett. {\bf 389}, 209 (2004).
\bibitem{lgaus2} K. Pachucki and J. Komasa, Phys. Rev. A {\bf 70}, 022513 (2004).
\bibitem{nak} T. Nakamura {\em et al.}, Phys. Rev. A {\bf 74}, 052503 (2006).
\bibitem{vecpol}  K. Pachucki, Phys. Rev. A {\bf 76}, 022508 (2007).
\bibitem{lihfs} K. Pachucki, Phys. Rev. A {\bf 66}, 062501 (2002).
\end{thebibliography}
\end{document}